\documentclass[a4paper,11pt]{article}
\usepackage{pos}
\usepackage{lineno}

\title{Charged-particle jet spectra in event-shape engineered Pb--Pb collisions at $\sqrt{s_{\rm NN}}$ = 5.02 TeV with ALICE}
 \ShortTitle{Charged jet spectra in event-shape engineered collisions with ALICE}

\author*[a]{Caitlin Beattie}

 \onbehalf{for the ALICE Collaboration}

\affiliation[a]{Yale University,\\
  272 Whitney Ave., New Haven, USA}

\emailAdd{caitlin.beattie@cern.ch}

\abstract{

The path-length dependence of jet quenching can help to constrain different jet quenching mechanisms in heavy-ion collisions. However, measuring an explicit value for this dependence has proven challenging. Traditional approaches, which consider anisotropic jet suppression arising from geometric asymmetries, have successfully measured a non-zero azimuthal dependence of jet modification with respect to the event-plane angle of the collision. While such signals improve our qualitative understanding of this topic, extraction of an explicit dependence from these results is limited by fluctuations in the initial state and jet--medium interactions. A new approach to characterize the geometry of the collision is to use event-shape engineering, a technique that classifies events within a centrality class according to their elliptical anisotropies. By doing so, we gain an improved knowledge of the initial-state medium, consequently enabling better constraints on the average path length traversed by the jet. In these proceedings, new results of jet spectra from event-shape-engineered collisions at ALICE will be presented.

}

\FullConference{%
  11th International Conference on Hard and Electromagnetic Probes of High-Energy Nuclear Collisions\\
  26-31 March 2023\\
  Aschaffenburg, Germany
}

\begin{document}
\maketitle

\section{Introduction}
Establishing experimental constraints of the path-length dependence of jet energy loss has proven to be an elusive goal. Because jets are internally generated probes of a collision, information of their spatial origin is not available, thus prohibiting the determination of the length of medium that a given jet traverses. Studies of the path-length dependence of jet energy loss must therefore take a statistical approach. \par

Traditionally, such studies can be classified into two broad groups: correlation analyses and azimuthal studies. Correlation analyses isolate some trigger object (for example, a jet or a high-$p_{\rm T}$ hadron) and consider the correlations between other objects in an event and this trigger. Alternatively, azimuthal studies consider the spectra of objects of interest as a function of their azimuthal difference from the event plane. While such measurements have consistently shown signals compatible with path-length dependent quenching \cite{ALICEjetv2, ALICEjetparticlev2, ATLASjetv2}, it has been difficult to extract explicit constraints from these results alone. Interpretations of correlation studies are often limited by fluctuations in jet--medium interactions, which are particularly obfuscating when occurring simultaneously in a trigger and its recoiling particles \cite{Zapp}. Interpretations of azimuthal studies are restricted by fluctuations in the initial-state medium, which preclude identification of the paths that jets traverse through the medium.\par

\begin{figure}[h!]
    \begin{center}
    \includegraphics[width = 0.925\textwidth]{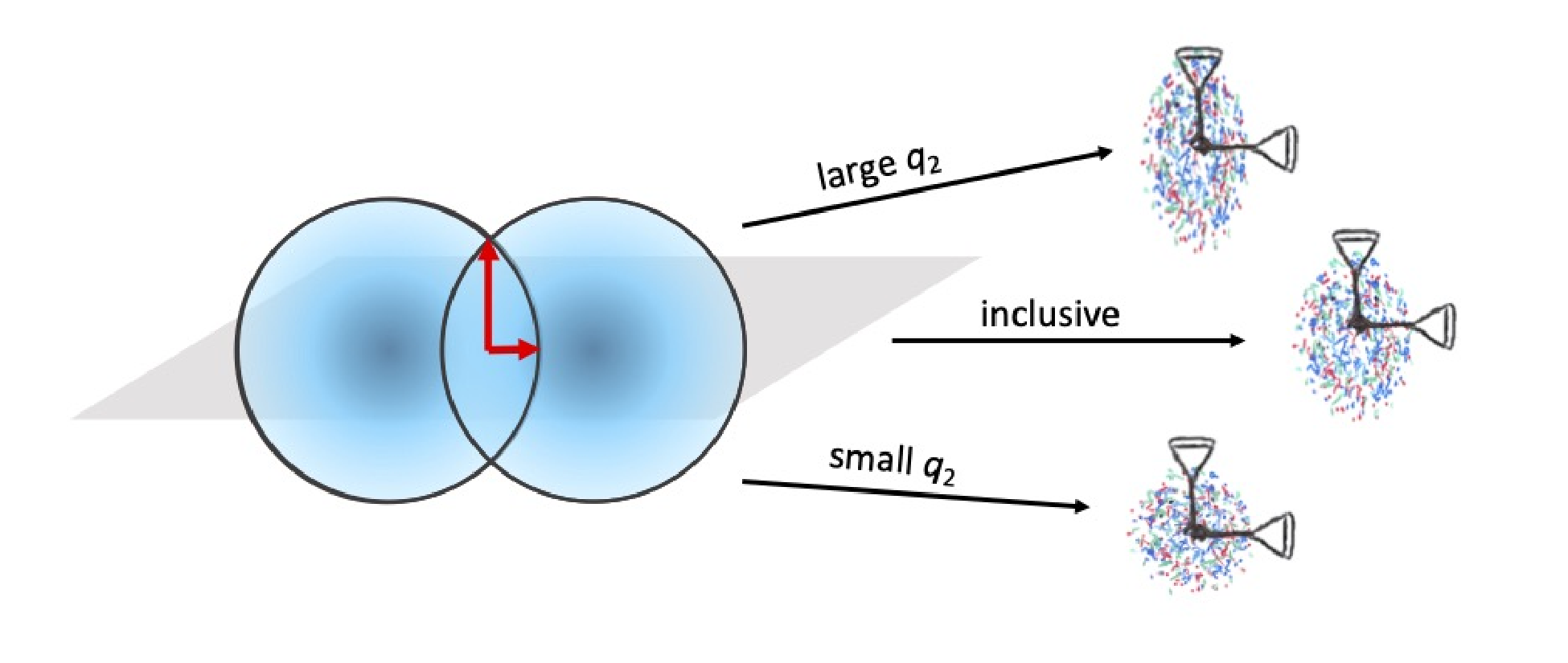}
    \end{center}
    \caption{A schematic of how event-shape engineering can be used to constrain the QGP shape and study the path-length dependence of jet energy loss. Here, $q_{\rm 2}$ represents the magnitude of the second-order flow harmonic. Large $q_{\rm 2}$ values correspond to more elliptical events, while small $q_{\rm 2}$ values correspond to more spherical events. Figure taken from Ref.~\cite{BeattieThesis}.}
    \label{fig:ESEjets}
\end{figure}

In these proceedings, a novel technique is presented to address these limitations. Event-shape engineering (ESE) is an analysis technique that exploits the fact that there exists a significant distribution of event geometries within any given centrality class \cite{ESE}. By classifying events within a centrality class (and thus with similar thermodynamic properties) according to their ellipticities using the quantity $q_{\rm 2} = |\textbf{Q}_{\rm 2}|/\sqrt{M}$ (where \textbf{Q}$_{\rm 2}$ is defined in Eq.~\ref{eq:bigQ2}), ESE improves upon existing azimuthal measurements by maximizing (or minimizing) the differences between in-plane and out-of-plane pathlengths \cite{trajectumESE}. A schematic of this idea, and how it can be used to study jet energy loss, can be seen in Fig.~\ref{fig:ESEjets}.

\section{Experimental Details}

\begin{figure}[h!]
    \begin{center}
    \includegraphics[width = 0.4\textwidth]{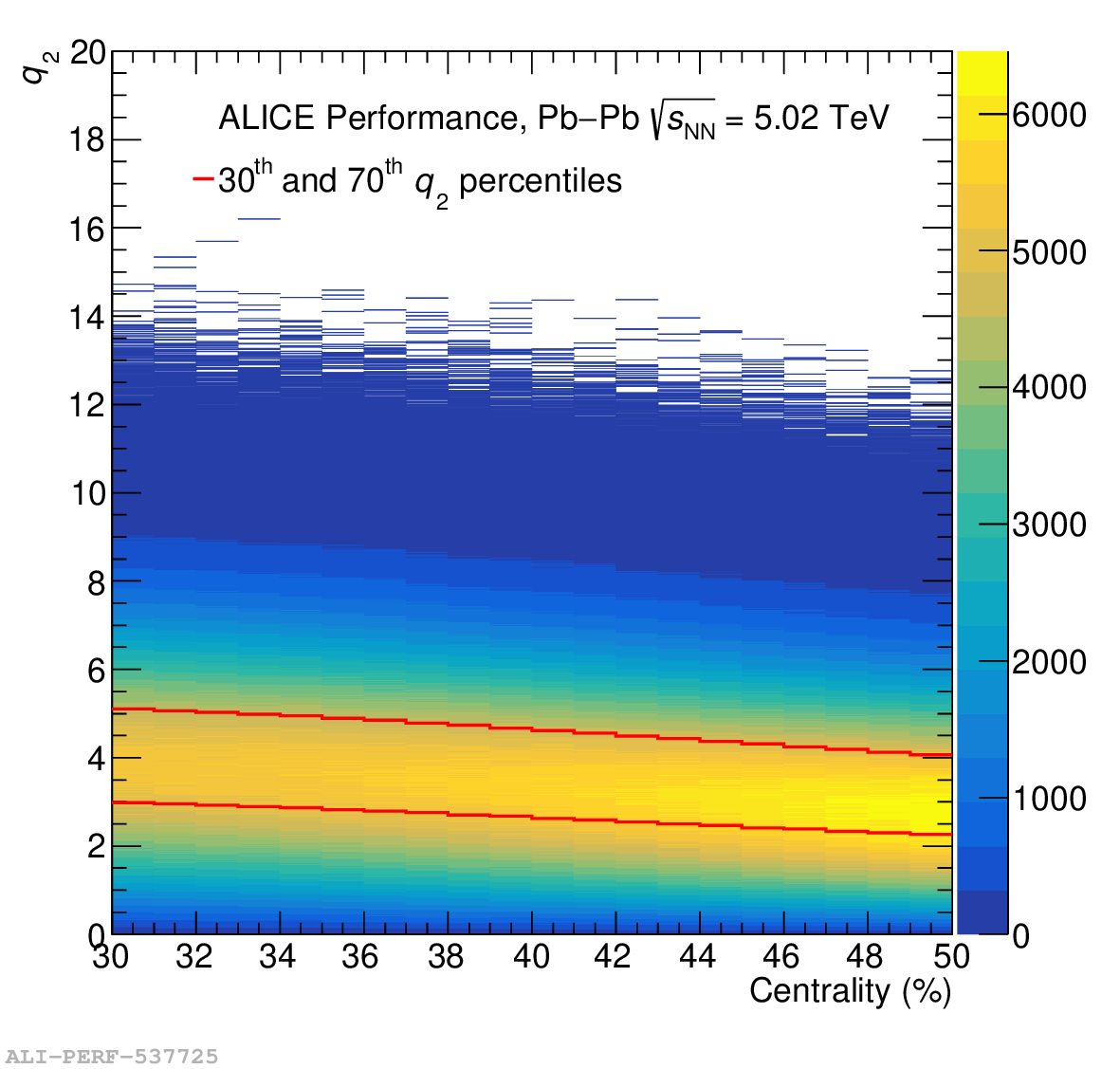}
    \end{center}
    \caption{Distribution of $q_{\rm 2}$ values as a function of event centrality.}
    \label{fig:q2}
\end{figure}

This analysis was performed using data from semicentral (30--50\%) Pb--Pb collisions taken with the ALICE detector at $\sqrt{s_{\rm NN}} = 5.02$ TeV \cite{ALICE}. Charged-particle jets were measured using tracks from the Inner Tracking System and Time Projection Chamber ($-0.9 < \eta < $ 0.9). These jets were reconstructed with the anti-$k_{\rm T}$ algorithm and $p_{\rm T}$-recombination scheme, using resolution parameters $R =$ 0.2 and 0.4 \cite{Fastjet}. Jets were required to have a leading track with 5 GeV/\textit{c} $ < p_{\rm T} < $ 100 GeV/\textit{c} and pseudorapidity within $|\eta_{\rm jet}| < 0.9 - R$. The transverse momentum of each jet was corrected using a pedestal-style, area-based subtraction method \cite{background}. The event-plane $\Psi_{\rm 2}$ was determined using the V0A detector. The event ellipticity was constrained using the magnitude of the second-order flow harmonic $q_{\rm 2} = |\textbf{Q}_{\rm 2}|/\sqrt{M}$, where $M$ is the charged-particle multiplicity and

\begin{equation}
        \textbf{Q}_{2} = \left( \sum_{i} w_{i} \cos(2\varphi_{i}), \sum_{i} w_{i} \sin(2\varphi_{i}) \right).
        \label{eq:bigQ2}
\end{equation}
Here, $\varphi_{i}$ are the azimuthal angles of the charged particles and $w_{\rm i}$ are the signal weights. The $q_{\rm 2}$ distribution was measured using the V0C and is shown in Fig.~\ref{fig:q2}. Separate detectors were used for the determination of $q_{\rm 2}$ and $\Psi_{\rm 2}$ to reduce contributions from auto-correlations.

 Jets were then classified according to their event-plane angle $\Delta\varphi = \Psi_{\rm 2} -\varphi$, as well as the shape of the event from which they were generated. The differentiated jet spectra were then unfolded for $p_{\rm T}$ and $\Delta\varphi$ smearing using a 2D Bayesian procedure \cite{Bayes, RooUnfold}. The ratios of out-of-plane to in-plane jet spectra were computed, accounting for correlated systematics. The three-sub-event method was then used to correct these ratios (and the corresponding spectra) for the second-order event-plane resolution \cite{threeSubEvent}. The results of this analysis procedure are included in the section below.

\section{Results}
\begin{figure}[h!]
    \begin{center}
    \includegraphics[width = 0.4\textwidth]{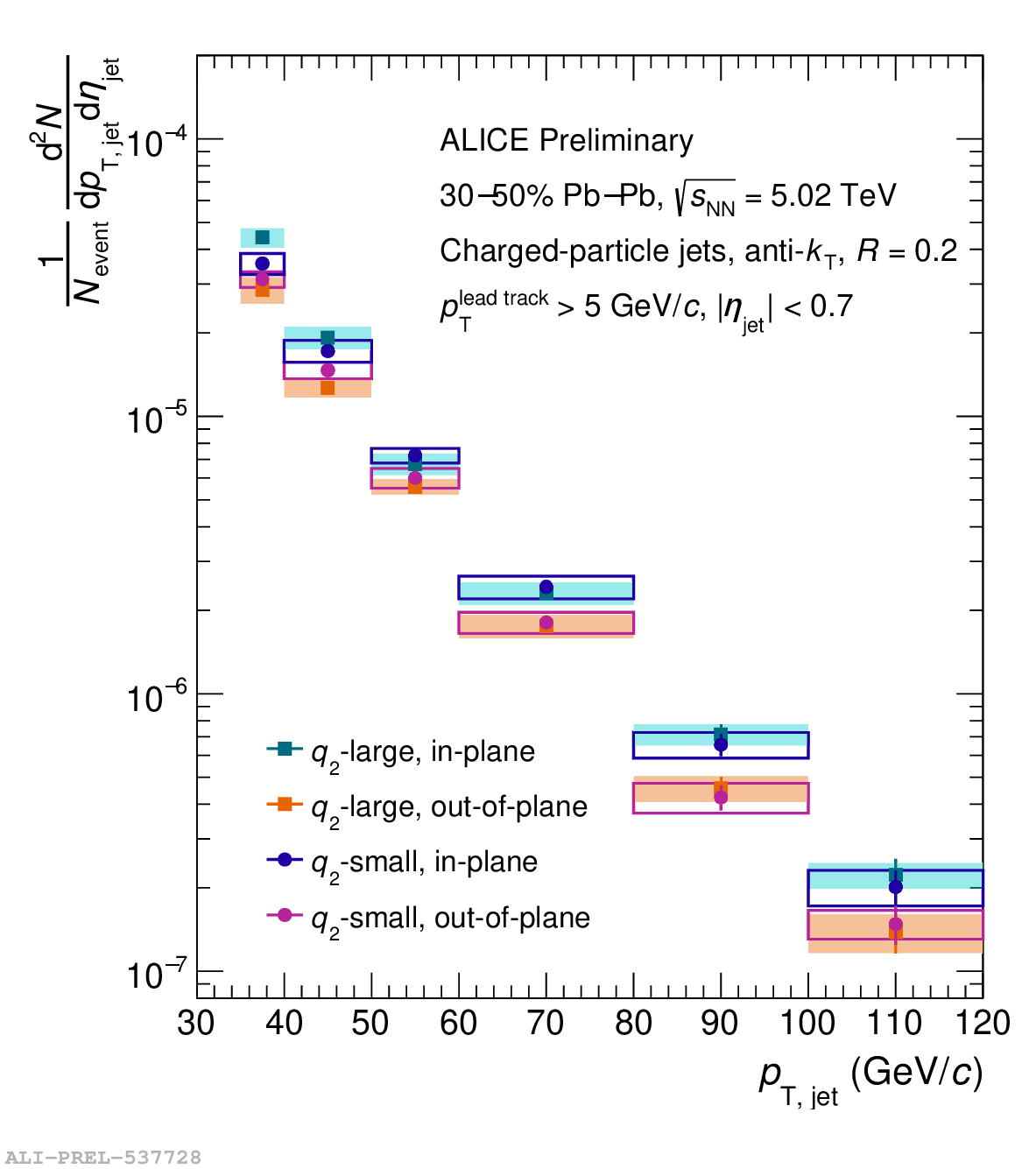}
     \includegraphics[width = 0.4\textwidth]{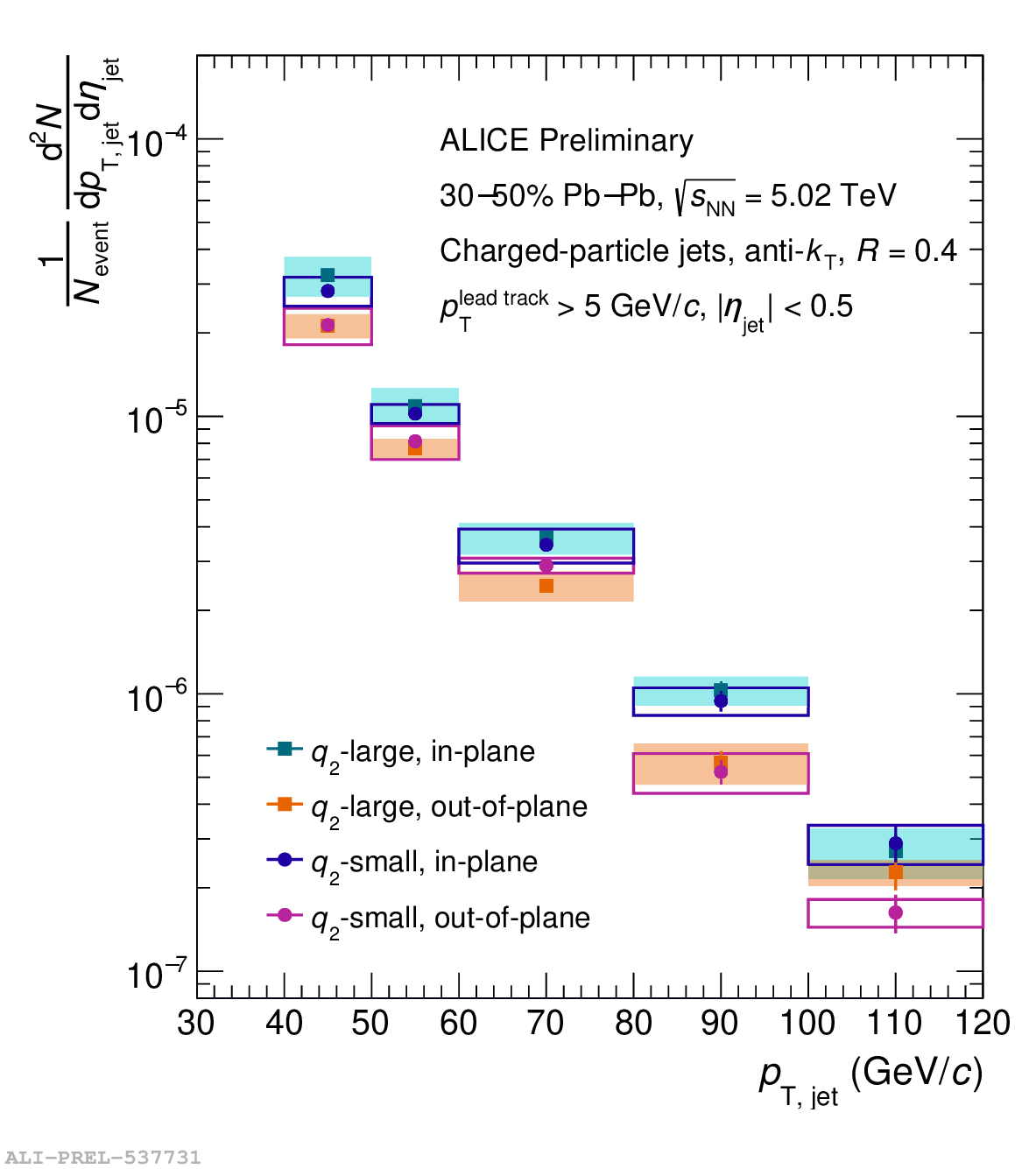}
    \end{center}
    \caption{Jet yields for $R = 0.2$ (left) and $R = 0.4$ (right) jets, separated by event-plane angle and event ellipticity.}
    \label{fig:yields}
\end{figure}

Jet spectra classified according to the event-shape and event-plane angle are shown in Fig.~\ref{fig:yields}. An apparent separation between in-plane and out-of plane jets can be seen in these spectra. However, the systematic uncertainties are large, and it is difficult to make quantitative claims from these results alone. The ratios of these spectra are therefore useful for understanding the magnitude of the difference between in-plane and out-of-plane jet yields, as well as any difference arising from event shapes. \par

\begin{figure}[h!]
    \begin{center}
    \includegraphics[width = 0.59\textwidth]{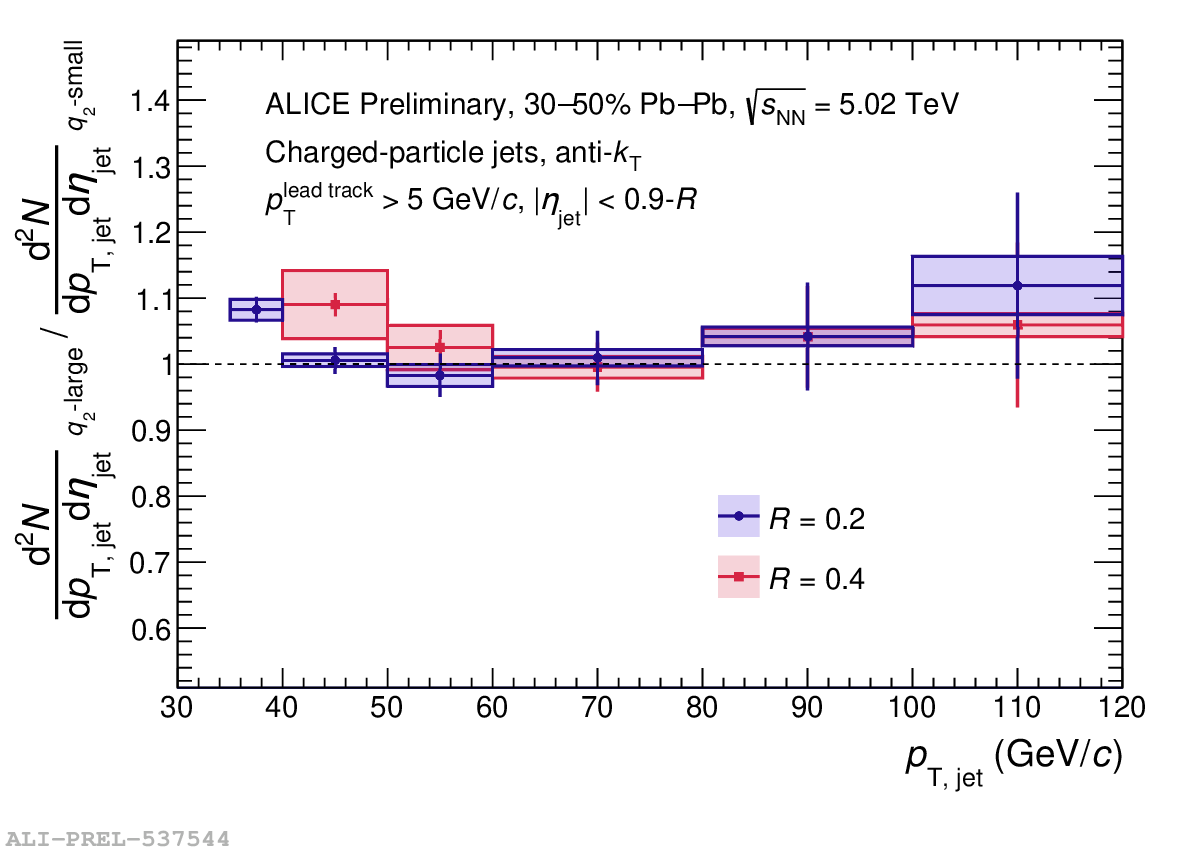}
    \end{center}
    \caption{Ratio of azimuthally integrated jet yields from elliptical events to round events.}
    \label{fig:flatness}
\end{figure}

Ratios of the jet yields from different event classes can be seen in Fig.~\ref{fig:flatness}. It is observed that these ratios are consistent with unity for $R = 0.2$ and $R = 0.4$ jets. This result demonstrates that azimuthally-integrated jet production is not sensitive to the shape of the underlying event as determined at forward rapidity using the V0C. This is in contrast with soft particle spectra measured at midrapidity, where a yield enhancement is seen in elliptical events \cite{ESE276}. Jet yields, on the other hand, are primarily modified by quenching effects. They are thus less sensitive to e.g. the blast wave expansion of the medium.

\begin{figure}[h!]
    \begin{center}
    \includegraphics[width = 0.48\textwidth]{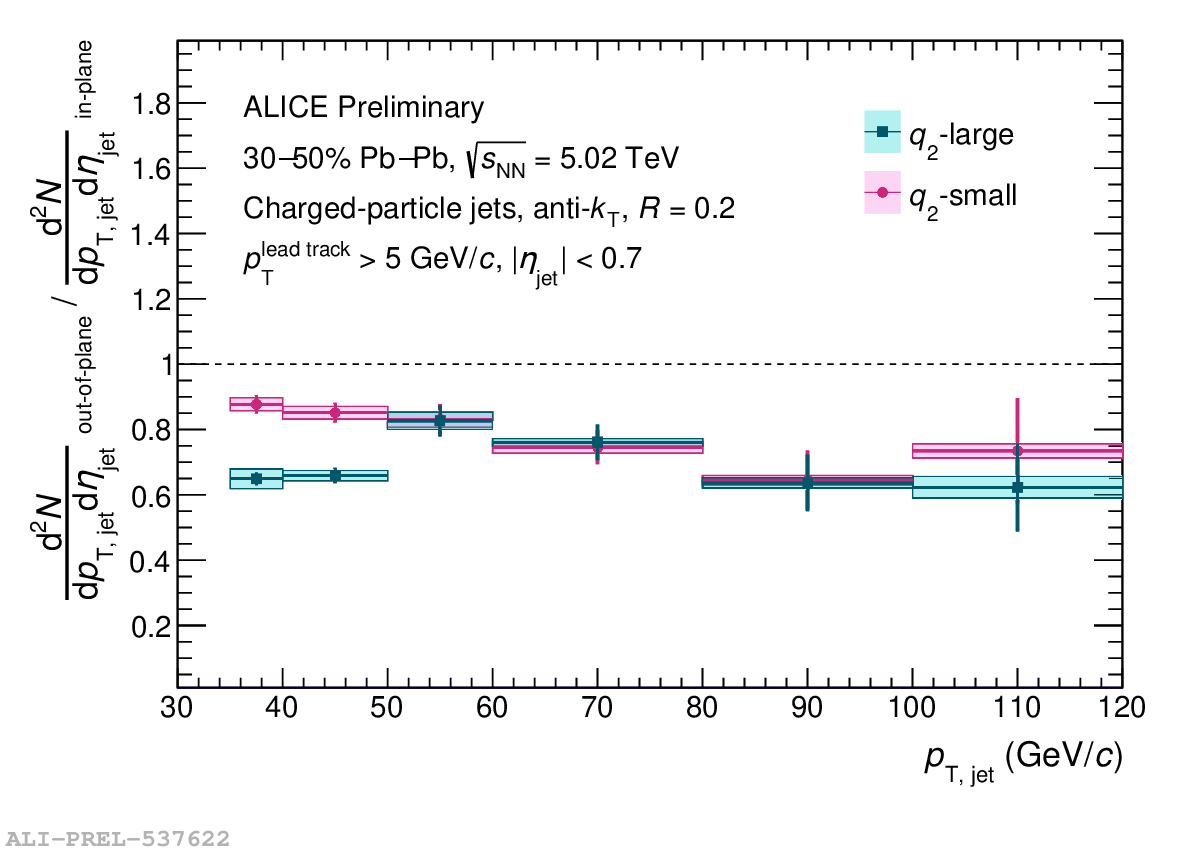}
    \includegraphics[width = 0.48\textwidth]{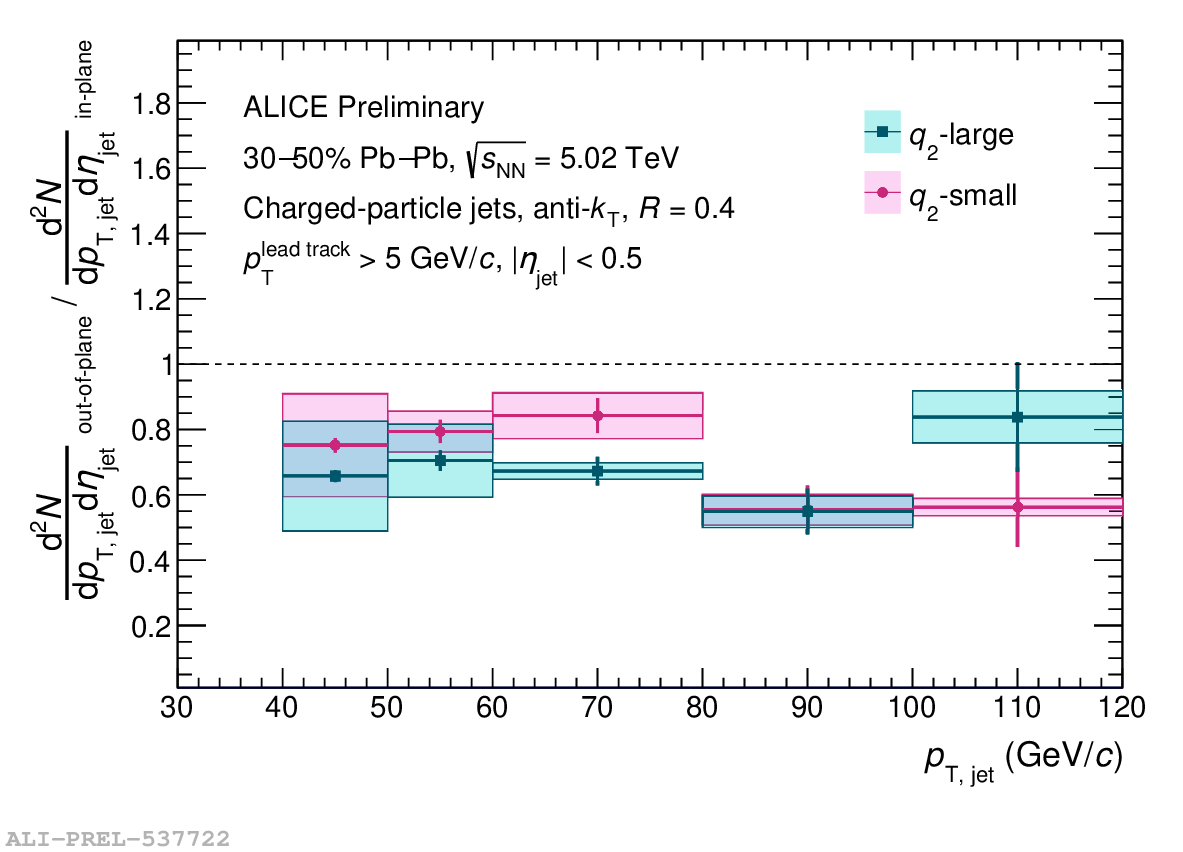}
    \end{center}
    \caption{Ratios of out-of-plane to in-plane jet yields for $R = 0.2$ (left) and $R = 0.4$ (right). Different colors represent different event shape classifications.}
    \label{fig:signal}
\end{figure}

Finally, the ratios of out-of-plane to in-plane jet yields are shown in Fig.~\ref{fig:signal}. These ratios are observed to be below unity for the full $p_{\rm T}$ range, consistent with the non-zero jet $v_{\rm 2}$ measurement \cite{ALICEjetv2}. Further conclusions about the event-shape dependence of this measurement are precluded for the $R = 0.4$ result due to large uncertainties that arise from sensitivity to the combinatorial background. However, an additional ellipticity-dependent signal is observed for $R = 0.2$ jets with $p_{\rm T} <$ 50 GeV/\textit{c}. Here, it is seen that the ratio of out-of-plane to in-plane jets is more suppressed in more elliptical events than in more round events. This result is compatible with the notion that the relative suppression of out-of-plane to in-plane jets should correspond with the relative difference between out-of-plane and in-plane path lengths.

\section{Conclusions}
Promising path-length dependent signals of jet suppression have been observed using event-shape engineering. However, model calculations are required to extract explicit constraints from these experimental measurements. With the development of jet quenching models that incorporate event-by-event hydrodynamic fluctuations, such comparisons are becoming possible \cite{LBT}, but require significant computational power that has not yet been accessible. Additionally, the large volume of data that will become available during the LHC Run 3 will enable higher precision measurements in the high-$p_{\rm T}$ regime. The combination of increased data volume and the availability of phenomenological models will thus open new routes for the usage of ESE to study and interpret jet production and the path-length dependence of energy loss.


\begin{thebibliography}{99}

\bibitem{ALICEjetv2}
\textbf{ALICE} Collaboration, J. Adam et al., “Azimuthal anisotropy of charged jet production in $\sqrt{s_{\rm NN}} =$ 2.76 TeV Pb--Pb collisions”, \textit{Phys. Lett. B} \textbf{753} (2016) 511--525, arXiv:1509.07334 [nucl-ex].

\bibitem{ALICEjetparticlev2}
\textbf{ALICE} Collaboration, “Azimuthal anisotropy of jet particles in p--Pb and Pb--Pb collisions at $\sqrt{s_{\rm NN}} =$ 5.02 TeV”, arXiv:2212.12609 [nucl-ex].

\bibitem{ATLASjetv2}
\textbf{ATLAS} Collaboration, G. Aad et al., “Measurement of the Azimuthal Angle Dependence of Inclusive Jet Yields in Pb+Pb Collisions at $\sqrt{s_{\rm NN}} = 2.76$ TeV with the ATLAS detector”, \textit{Phys. Rev. Lett.} \textbf{111} (2013) 152301, arXiv:1306.6469 [hep-ex].

\bibitem{Zapp}
J. G. Milhano and K. C. Zapp, “Origins of the di-jet asymmetry in heavy ion collisions”, \textit{Eur. Phys. J. C} \textbf{76} (2016) 288, arXiv:1512.08107 [hep-ph].

\bibitem{ESE}
J. Schukraft, A. Timmins, and S. A. Voloshin, “Ultra-relativistic nuclear collisions: event shape engineering”, \textit{Phys. Lett. B} \textbf{719} (2013) 394--398, arXiv:1208.4563 [nucl-ex].

\bibitem{trajectumESE}
C. Beattie, G. Nijs, M. Sas, and W. van der Schee, “Hard probe path lengths and event-shape engineering of the quark--gluon plasma”, \textit{Phys. Lett. B} \textbf{836} (2023) 137596, arXiv:2203.13265 [nucl-th].

\bibitem{BeattieThesis}
C. Beattie, \textit{Pathlength-dependent jet quenching in the quark--gluon plasma at ALICE}, Ph.D. thesis,
Yale U., 2023, CERN-THESIS-2023-030.

\bibitem{ALICE}
\textbf{ALICE} Collaboration, B. Abelev et al., “Performance of the ALICE Experiment at the CERN LHC”, \textit{Int. J. Mod. Phys. A} \textbf{29} (2014) 1430044, arXiv:1402.4476 [nucl-ex].

\bibitem{Fastjet}
M. Cacciari, G. P. Salam, and G. Soyez, “FastJet User Manual”, \textit{Eur. Phys. J. C} \textbf{72} (2012) 1896,
arXiv:1111.6097 [hep-ph].

\bibitem{background}
M. Cacciari and G. P. Salam, “Pileup subtraction using jet areas”, \textit{Phys. Lett. B} \textbf{659} (2008)
119-–126, arXiv:0707.1378 [hep-ph].

\bibitem{Bayes}
G. D’Agostini, “A multidimensional unfolding method based on Bayes’ theorem”, \textit{Nucl. Instrum. Meth. A} \textbf{362} (1995) 487–498.

\bibitem{RooUnfold}
T. Adye, “Unfolding algorithms and tests using RooUnfold”, \textit{Proceedings of the PHYSTAT 2011 Workshop}, 313–318. CERN, Geneva, 2011, arXiv:1105.1160 [physics.data-an].

\bibitem{threeSubEvent}
A. Poskanzer and S. Voloshin, “Methods for analyzing anisotropic flow in relativistic nuclear collisions”, \textit{Phys. Rev. C} \textbf{58} (1998) 1671–1678.

\bibitem{ESE276}
\textbf{ALICE} Collaboration, J. Adam \textit{et al.}, “Event shape engineering for inclusive spectra and elliptic flow in Pb–Pb collisions at $\sqrt{s_{\rm NN}} = 2.76$ TeV”, \textit{Phys. Rev. C} \textbf{93} (2016) 034916, arXiv:1507.06194 [nucl-ex].

\bibitem{LBT}
Y. He, T. Luo, X.-N. Wang, and Y. Zhu, “Linear Boltzmann Transport for Jet Propagation in the
Quark--Gluon Plasma: Elastic Processes and Medium Recoil”, \textit{Phys. Rev. C} \textbf{91} (2015) 054908,
arXiv:1503.03313 [nucl-th]. [Erratum: Phys. Rev. C 97, 019902 (2018)].


\end{thebibliography}
\end{document}